\documentclass[10pt,conference]{IEEEtran}
\usepackage{amsfonts}
\usepackage{amsfonts}
\usepackage{setspace}
\usepackage{clrscode}

\usepackage{amsmath}
\usepackage{amssymb}
\usepackage[dvips]{graphicx}
\usepackage{epsfig}
\usepackage{hyperref}

\usepackage{amsmath}
\usepackage{amssymb}
\usepackage[dvips]{graphicx}
\usepackage{epsfig}

\newcommand{\tsnr}{{\text{\footnotesize{SNR}}}}

\newcommand{\E}{\mathbb{E}}

\newcommand{\C}{{\sf{C}}}

\newcommand{\Pb}{\bar{P}}

\newcommand{\figsize}{0.4}

\newtheorem{Lem1}{Proposition}
\newtheorem{Lem2}{Lemma}
\newtheorem{Lem}{Theorem}

\begin{document}

%
\title{Secure Broadcasting over Fading Channels with Statistical QoS Constraints}



%
\author{\authorblockN{Deli Qiao, Mustafa Cenk Gursoy, and Senem
Velipasalar}
\authorblockA{Department of Electrical Engineering\\
University of Nebraska-Lincoln, Lincoln, NE 68588\\ Email:
dqiao726@huskers.unl.edu, gursoy@engr.unl.edu,
velipasa@engr.unl.edu} }


\maketitle

\begin{abstract}\footnote{This work was supported by the National Science Foundation under Grants CNS--0834753, and CCF--0917265.}
In this paper, the fading broadcast channel with confidential
messages is studied in the presence of statistical quality of
service (QoS) constraints in the form of limitations on the buffer
length. We employ the effective capacity formulation 
to measure the throughput of the confidential and common messages. We
assume that the channel side information (CSI) is available at both the
transmitter and the receivers. Assuming average power
constraints at the transmitter side, we first define the
\emph{effective secure throughput region}, and prove that the
throughput region is convex. Then, we obtain the optimal power
control policies that achieve the boundary points of the
\emph{effective secure throughput region}.
\end{abstract}

\section{Introduction}
Security is an important issue in wireless systems due to the
broadcast nature of wireless transmissions. In a pioneering work,
Wyner in \cite{wyner} addressed the security problem from an
information-theoretic point of view and considered a wire-tap
channel model. He proved that secure transmission of confidential
messages to a destination in the presence of a degraded wire-tapper
can be achieved, and he established the secrecy capacity which is
defined as the highest rate of reliable communication from the
transmitter to the legitimate receiver while keeping the wire-tapper
completely ignorant of the transmitted messages. Recently, there has
been numerous studies addressing information theoretic security. For
instance, the impact of fading has been investigated in \cite{lai},
where it has been shown that a non-zero secrecy capacity can be
achieved even when the eavesdropper channel is better than the main
channel on average. The secrecy capacity region of the fading
broadcast channel with confidential messages and associated optimal
power control policies have been identified in \cite{liangsecure},
where it is shown that the transmitter allocates more power as the
strength of the main channel increases with respect to that of the
eavesdropper channel.

In addition to security issues, providing acceptable performance and
quality is vital to many applications. For instance, voice over IP
(VoIP) and interactive-video (e.g,. videoconferencing) systems are
required to satisfy certain buffer or delay constraints. In this
paper, we consider statistical QoS constraints in the form of
limitations on the buffer length, and incorporate the concept of
effective capacity \cite{dapeng}, which can be seen as the maximum
constant arrival rate that a given time-varying service process can
support while satisfying statistical QoS guarantees. The analysis
and application of effective capacity in various settings have
attracted much interest recently (see e.g.,
\cite{jia}--\cite{liu-cooperation} and references therein). We
define the \emph{effective secrecy throughput region} as the maximum
constant arrival rate pairs that can be supported while the service
rate is confined by the secrecy capacity region. We assume that the
channel side information is known at both the transmitter and receivers.
Then, following a similar analysis as shown in \cite{liangsecure}, we
obtain the optimal power allocation policies that achieve
points on the boundary of the effective secrecy throughput region.

The rest of the paper is organized as follows. Section II briefly
describes the system model and the necessary preliminaries on
statistical QoS constraints and effective capacity. In Section III,
we present our main results on the optimal power control policies.
Finally, Section IV concludes the paper.

\section{System Model and Preliminaries}

\subsection{System Model}

\begin{figure}
\begin{center}
\includegraphics[width=0.35\textwidth]{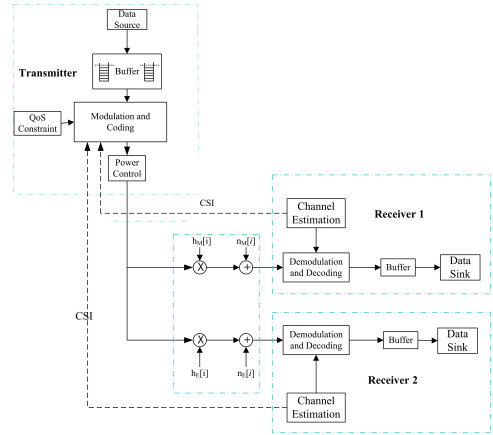}
\caption{The general system model.}\label{fig:systemmodel}
\end{center}
\end{figure}
We consider a scenario in which a single transmitter broadcasts messages to two receivers. The transmitter wishes to send receiver 1 confidential messages that need to kept secret from receiver 2, and also at the same time send common messages to both receivers. A depiction of the system model is given in Figure \ref{fig:systemmodel}. It is assumed that the transmitter generates data sequences which
are divided into frames of duration $T$. These data frames are
initially stored in the buffer before they are transmitted over the
wireless channel. The channel input-output relationships are given by
\begin{align}
Y_1[i]=h_1[i]X[i]+Z_1[i]\,\text{and}\,Y_2[i]=h_2[i]X[i]+Z_2[i]
\end{align}
where $i$ is the frame index, $X[i]$ is the channel input in the
$i$th frame, and $Y_1[i]$ and $Y_2[i]$ represent the channel outputs
at the receivers 1 and 2 at frame $i$, respectively. We assume that
$\{h_j[i],\, j=1,2\}$'s are jointly stationary and ergodic
discrete-time processes, and we denote the magnitude-square of the
fading coefficients by $z_j[i]=|h_j[i]|^2$. Considering that
receiver 1 is the main user to which we send both the common and confidential messages, while receiver 2, to which we send only the common messages, can be regarded as an eavesdropper for the confidential messages,  we replace $z_1$ with $z_M$ and $z_2$ with
$z_E$ to increase the clarity in the subsequent formulations.  The channel input is subject
to an average power constraint $\E\{|X[i]|^2\} \le \Pb$, and we
assume that the bandwidth available for the system is $B$. Above,
$Z_j[i]$ is a zero-mean, circularly symmetric, complex Gaussian
random variable with variance $\E\{|Z_j[i]|^2\} = N_j$. The additive
Gaussian noise samples $\{Z_j[i]\}$ are assumed to form an
independent and identically distributed (i.i.d.) sequence.

Note that we denote the average transmitted signal-to-noise ratio
with respect to receiver 1 as $\tsnr=\frac{\Pb}{N_1 B}$. We also denote
$P[i]$ as the instantaneous transmit power in the $i$th frame. Now, the
instantaneous transmitted SNR level for receiver 1 becomes
$\mu^1[i]=\frac{P[i]}{N_1 B}$. Then, the average power constraint at
the base station is equivalent to the average SNR constraint
$\E\{\mu^1[i]\}\le \tsnr$ for receiver 1. If we denote the ratio
between the noise powers of the two channels as
$\gamma=\frac{N_1}{N_2}$, the instantaneous transmitted SNR level
for receiver 2 becomes $\mu^2[i]=\gamma\mu^1[i]$.

We denote $\mathbf{z}=(z_E,z_M)$ as the vector composed of the
channel states for receivers 2 and 1. Under the block fading
assumption, the maximal instantaneous service rate is decided by the
secrecy capacity region of the Gaussian broadcast channel with confidential messages (BCC) for each interval with a
pair of specific channel states realizations $(z_E,z_M)$. We define
$\mu=(\mu_0(\mathbf{z}),\mu_1(\mathbf{z}))$ as the power allocation
policies for the common and confidential messages, respectively. We
denote the region $\mathcal{Z}=\left\{\mathbf{z}: \, z_M>\gamma
z_E\right\}$ as the region in which both confidential and common messages are transmitted.
$\mathcal{Z}^c$ represents the region where $ z_M\le\gamma z_E$. If $\mathbf{z}$ lies in $\mathcal{Z}^c$, only the common messages are transmitted.
Note that $\mu_1(\mathbf{z})=0$ for $\mathbf{z}\in\mathcal{Z}^c$. We then
define the set $\mathcal{U}$ as the power allocation policies
satisfying the power constraints
\begin{align}\label{eq:avgpower}
\hspace{-.5cm}\mathcal{U}=\left\{\mu:\, \E_{\mathbf{z}\in\mathcal{Z}
}\{\mu_0(\mathbf{z})+\mu_1(\mathbf{z})\}+\E_{\mathbf{z}\in\mathcal{Z}^c}\{\mu_0(\mathbf{z})\}\le\tsnr\right\}.
\end{align}

With the above notations, the secrecy capacity region of the fading
BCC is given by \cite{liangsecure}
\begin{small}
\begin{align}\label{eq:cregion}
&\mathcal{R}_s=\bigcup_{\mu\in\mathcal{U}}\nonumber\\
&\left\{
\begin{array}{l}
(R_0,R_1):\\
R_0\le\min\left\{
\begin{array}{l}
\E_{\mathbf{z}\in\mathcal{Z}}\left\{\log_2\left(1+\frac{\mu_0(\mathbf{z})z_M}{1+
\mu_1(\mathbf{z})z_M}\right)\right\}\\
+\E_{\mathbf{z}\in\mathcal{Z}^c}\left\{\log_2\left(1+\mu_0(\mathbf{z})z_M\right)\right\},\\
\E_{\mathbf{z}\in\mathcal{Z}}\left\{\log_2\left(1+\frac{\gamma\mu_0(\mathbf{z})z_E}{1+
\gamma\mu_1(\mathbf{z})z_E}\right)\right\}\\
+\E_{\mathbf{z}\in\mathcal{Z}^c}\left\{\log_2\left(1+\gamma\mu_0(\mathbf{z})z_E\right)\right\}
\end{array}
\right\}\\
R_1\le\E_{\mathbf{z}\in\mathcal{Z}}\left\{\log_2\left(1+\mu_1(\mathbf{z})z_M\right)-\log_2\left(1+\gamma\mu_1(\mathbf{z})z_E\right)\right\}
\end{array}
\right\}
\end{align}
\end{small}

\subsection{Statistical QoS Constraints and Effective Secure Throughput}
In \cite{dapeng}, Wu and Negi defined the effective capacity as the
maximum constant arrival rate\footnote{For time-varying arrival
rates, effective capacity specifies the effective bandwidth of the
arrival process that can be supported by the channel.} that a given
service process can support in order to guarantee a statistical QoS
requirement specified by the QoS exponent $\theta$. If we define $Q$
as the stationary queue length, then $\theta$ is the decay rate of
the tail distribution of the queue length $Q$:
\begin{equation}
\lim_{q \to \infty} \frac{\log P(Q \ge q)}{q} = -\theta.
\end{equation}
Therefore, for large $q_{\max}$, we have the following approximation
for the buffer violation probability: $P(Q \ge q_{\max}) \approx
e^{-\theta q_{\max}}$. Hence, while larger $\theta$ corresponds to
more strict QoS constraints, smaller $\theta$ implies looser QoS
guarantees. Similarly, if $D$ denotes the steady-state delay
experienced in the buffer, then $P(D \ge d_{\max}) \approx
e^{-\theta \delta d_{\max}}$ for large $d_{\max}$, where $\delta$ is
determined by the arrival and service processes
\cite{tangzhangcross2}.

The effective capacity is given by
\begin{align}\label{eq:effectivedefi}
C(\theta)=-\frac{\Lambda(-\theta)}{\theta}=-\lim_{t\rightarrow\infty}\frac{1}{\theta
t}\log_e{\mathbb{E}\{e^{-\theta S[t]}\}} \,\quad \text{bits/s},
\end{align}
where the expectation is with respect to $S[t] =
\sum_{i=1}^{t}s[i]$, which is the time-accumulated service process.
$\{s[i], i=1,2,\ldots\}$ denote the discrete-time stationary and
ergodic stochastic service process. We define the effective capacity
region obtained when the service rate is confined by the secrecy
capacity region as the \emph{effective secrecy throughput region}.

In this paper, in order to simplify the analysis while considering
general fading distributions, we assume that the fading coefficients
stay constant over the frame duration $T$ and vary independently for
each frame and each user. In this scenario, $s[i]=T R[i]$, where
$R[i]$ is the instantaneous service rate for common or confidential
messages in the $i$th frame duration $[iT;(i+1)T]$. Then,
(\ref{eq:effectivedefi}) can be written as
\begin{align}
C(\theta)&=
-\frac{1}{\theta T}\log_e\mathbb{E}_{\mathbf{z}}\{e^{-\theta T
R[i]}\}\,\quad \text{bits/s}, \label{eq:effectivedefirate}
\end{align}
where $R[i]$ denotes the instantaneous rate sequence with respect to
$\mathbf{z}$. (\ref{eq:effectivedefirate}) is obtained using the
fact that instantaneous rates $\{R[i]\}$ vary independently.
The \emph{effective secrecy throughput} normalized by bandwidth $B$
is
\begin{equation}\label{eq:normeffectivedefi}
\C(\theta)=\frac{C(\theta)}{B} \quad \text{bits/s/Hz}.
\end{equation}

\section{Effective Secrecy Throughput Region}

In this section, we investigate the fading broadcast channel with
confidential message (BCC) by incorporating the statistical QoS
constraints. Liang \emph{et al.} in \cite{liangsecure} have shown
that the fading channel can be viewed as a set of parallel
subchannels with each corresponding to one fading state. In \cite{liangsecure}, it has been assumed that no delay constraints are imposed on
the transmitted messages. Under such an assumption, the ergodic secrecy capacity region is determined and the optimal power
allocation policies achieving the boundary of the capacity region are identified.

In this paper, we analyze the performance under statistical buffer constraints by considering the effective capacity formulation. According to the formula for effective capacity
(\ref{eq:effectivedefirate}), we first have the following result.
\begin{Lem1}
The effective secure throughput region of the fading BCC is
\begin{align}
\mathcal{C}_{es}&=\bigcup_{\mu\in\mathcal{U}}\Bigg\{(\C_0,\C_1):
\C_j\le -\frac{1}{\theta TB}\log_e\E\{e^{-\theta T
R_j[i]}\},\, \nonumber\\
&\hspace{3cm}\text{subject to }\, \forall \E\{\mathbf{R}\}\in
\mathcal{C}_s\Bigg\}
\end{align}
where $\mathbf{R}=(R_0,R_1)$ is the vector composed of the
instantaneous rates for the common and confidential messages, respectively.
\end{Lem1}

We assume that $\E\{\mathbf{R}\}$ can take any possible value
defined in the ergodic secrecy capacity region $\mathcal{C}_s$. Since the secrecy capacity region is convex \cite{liangsecure}, we can easily prove the following.
\begin{Lem}
The effective secrecy throughput region is convex.
\end{Lem}
\emph{Proof: }Let the two effective capacity pairs $\underline{\C}_1 = (\C_{10}, \C_{11})$ and $\underline{\C}_2 = (\C_{20}, \C_{21})$ belong to $\mathcal{C}_{es}$.
Therefore, there exists some $\mathbf{R}[i]$ and $\mathbf{R'}[i]$
for $\underline{\C}_1(\Theta)$ and $\underline{\C}_2(\Theta)$, respectively. By a time
sharing strategy, for any $\alpha\in(0,1)$, we know that
$\E\{\alpha\mathbf{R}[i]+(1-\alpha)\mathbf{R'}[i]\}\in\mathcal{R}_{s}$.
\begin{align}
&\alpha\underline{\C}_1+(1-\alpha)\underline{\C}_2\nonumber\\
&=-\frac{1}{\theta TB}\log_e\left(\E\left\{e^{-\theta
T\mathbf{R}[i]}\right\}\right)^\alpha\left(\E\left\{e^{-\theta
T\mathbf{R'}[i]}\right\}\right)^{1-\alpha}\nonumber\\
&=-\frac{1}{\theta TB}\log_e\left(\E\left\{\left(e^{-\theta
T\alpha\mathbf{R}[i]}\right)^{\frac{1}{\alpha}}\right\}\right)^\alpha\nonumber\\
&\phantom{-\frac{1}{\Theta TB}e^{-\Theta
T\mathbf{R'}}}\cdot\left(\E\left\{\left(e^{-\theta
T(1-\alpha)\mathbf{R'}[i]}\right)^{\frac{1}{1-\alpha}}\right\}\right)^{1-\alpha}\nonumber\\
&\leq-\frac{1}{\theta TB}\log_e\E\left\{e^{-\theta T\left(\alpha
\mathbf{R}[i]+(1-\alpha)\mathbf{R'}[i]\right)}\right\}.
\end{align}
Above, the vector operation is with respect to each component. The inequality in the last stage follows from
H\"{o}lder's inequality. Hence, $\alpha\underline{\C}_1+(1-\alpha)\underline{\C}_2$
still lies in the \emph{throughput region}. \hfill$\square$

Then, the points on the boundary surface of the effective throughput
region $(\C_0^*,\C_1^*)$ can be obtained by solving the following
optimization problem
\begin{align}\label{eq:optproblem}
\max_{\mu\in\mathcal{U}} \, \lambda_0\C_0+\lambda_1\C_1
\end{align}
where $\lambda=(\lambda_0,\lambda_1)$ is any vector in
$\mathfrak{R}^2_+$. The optimal power allocation policies for the
above problem can be solved using a similar approach as in
\cite{liangrelay}, which is stated in \cite[Lemma 2]{liangsecure} as
well. Note that due to the introduction of QoS constraints, the
service rate is limited by the channel conditions while the
maximization is over the effective throughput. We first have the
following result.

\begin{Lem2}
The optimal $\mu^*$ that solves (\ref{eq:optproblem}) falls into the
one of the following three cases:
\begin{align}
&\text{Case I: } R_{01}(\mu^*)< R_{02}(\mu^*)\, \text{and} \,
\mu^*\,\text{
maximizes} \, \lambda_0 \C_{01}(\mu)+\lambda_1\C_1(\mu);\nonumber\\
&\text{Case II: } R_{01}(\mu^*)>R_{02}(\mu^*)\,\text{and}\,\mu^*\,
\text{maximizes} \,\lambda_0 \C_{02}(\mu)+\lambda_1\C_1(\mu);\nonumber\\
&\text{Case III: } R_{01}(\mu^*)=R_{02}(\mu^*)\,\text{then}\,
\nonumber\\
&\hspace{0.8cm}A) \,\mu^* \,\text{maximizes}\, \lambda_0
\C_{01}(\mu)+\lambda_1\C_1(\mu),
\text{if}\, \C_{01}(\mu)>\C_{02}(\mu);\nonumber\\
&\hspace{0.8cm}B)\,\mu^* \,\text{maximizes}\, \lambda_0
\C_{02}(\mu)+\lambda_1\C_1(\mu),
\text{if}\,\C_{01}(\mu)<\C_{02}(\mu);\nonumber\\
&\hspace{0.8cm}C)\,\mu^*\,\text{maximizes}\, \lambda_0
\left(\delta\C_{01}(\mu)+(1-\delta)\C_{02}(\mu)\right)+\lambda_1\C_1(\mu),
\, \nonumber\\
&\hspace{0.5cm}\text{if there exists $0\le\delta\le1$ such that }
\C_{01}(\mu)=\C_{02}(\mu).
\end{align}
Above, $R_{01}(\mu)$ and $R_{02}(\mu)$ are the two terms that
$R_0(\mu)$ can take in the minimization in (\ref{eq:cregion}), and
$\C_{01}(\mu)$ and $\C_{02}(\mu)$ are the associated effective
throughput values.
\end{Lem2}

Next, we derive the optimal power allocation
$\mu^*$ that solves (\ref{eq:optproblem}) for the different cases detailed above. Note that the maximal confidential message rate is defined as
\begin{align}
R_1=\left\{\begin{array}{ll}
\log_2\left(\frac{1+\mu_1(\mathbf{z})z_M}{1+\gamma\mu_1(\mathbf{z})z_E}\right),\,&\mathbf{z}\in\mathcal{Z}\\
0,\,&\mathbf{z}\in\mathcal{Z}^c\end{array}\right.
\end{align}

\textbf{Case I}: The maximal instantaneous common message rate is
given by the rate of receiver 1
\begin{align}
R_{01}=\left\{\begin{array}{ll}
\log_2\left(1+\frac{\mu_0(\mathbf{z})z_M}{1+\mu_1(\mathbf{z})z_M}\right),\,&\mathbf{z}\in\mathcal{Z}\\
\log_2\left(1+\mu_0(\mathbf{z})z_M\right),\,&\mathbf{z}\in\mathcal{Z}^c\end{array}\right.
\end{align}
as long as the obtained power control policy satisfies
\begin{equation}\label{eq:case1cond}
R_{01}(\mu)<R_{02}(\mu).
\end{equation}

Then, the Lagrangian is given by
\begin{small}
\begin{align}
\mathcal{J}&=-\frac{\lambda_0}{\beta \log_e 2}\log_e
\Bigg(\int_{\mathbf{z}\in\mathcal{Z}}\left(1+\frac{\mu_0(\mathbf{z})z_M}{1+\mu_1(\mathbf{z})
z_M}\right)^{-\beta}p_{\mathbf{z}}(z_M,z_E)d\mathbf{z}\nonumber\\
&\hspace{1.5cm}+\int_{\mathbf{z}\in\mathcal{Z}^c}\left(1+\mu_0(\mathbf{z})z_M\right)^{-\beta}p_{\mathbf{z}}(z_M,z_E)d\mathbf{z}\Bigg)\nonumber\\
&-\frac{\lambda_1}{\beta \log_e
2}\log_e\Bigg(\int_{\mathbf{z}\in\mathcal{Z}}\left(\frac{1+\mu_1(\mathbf{z})z_M}{1+\gamma\mu_1(\mathbf{z})
z_M}\right)^{-\beta}p_{\mathbf{z}}(z_M,z_E)d\mathbf{z}\nonumber\\
&\hspace{1.5cm}+\int_{\mathbf{z}\in\mathcal{Z}^c}p_{\mathbf{z}}(z_M,z_E)d\mathbf{z}\Bigg)\nonumber\\
&\hspace{1cm}-\kappa\left(\E_{\mathbf{z}\in\mathcal{Z}
}\{\mu_0(\mathbf{z})+\mu_1(\mathbf{z})\}+\E_{\mathbf{z}\in\mathcal{Z}^c}\{\mu_0(\mathbf{z})\}\right)
\end{align}
\end{small}
With specific power control policies, the values of $(\C_0,\C_1)$
are determined. Hence, we can define the following
\begin{align}
\phi_0&=\int_{\mathbf{z}\in\mathcal{Z}}\left(1+\frac{\mu_0(\mathbf{z})z_M}{1+\mu_1(\mathbf{z})
z_M}\right)^{-\beta}p_{\mathbf{z}}(z_M,z_E)d\mathbf{z}\nonumber\\
&\hspace{1.5cm}+\int_{\mathbf{z}\in\mathcal{Z}^c}\left(1+\mu_0(\mathbf{z})z_M\right)^{-\beta}p_{\mathbf{z}}(z_M,z_E)d\mathbf{z}\\
\phi_1&=\int_{\mathbf{z}\in\mathcal{Z}}\left(\frac{1+\mu_1(\mathbf{z})z_M}{1+\gamma\mu_1(\mathbf{z})
z_E}\right)^{-\beta}p_{\mathbf{z}}(z_M,z_E)d\mathbf{z}\nonumber\\
&\hspace{1.5cm}+\int_{\mathbf{z}\in\mathcal{Z}^c}p_{\mathbf{z}}(z_M,z_E)d\mathbf{z}
\end{align}
Although implicitly, $\phi_0$ and $\phi_1$ in the other cases in the
following analysis can be defined similarly. These two equations can
be viewed as additional constraints that the power control policies
need to satisfy, i.e., the right hand side (RHS) of the two
equations are also functions of $(\phi_0,\phi_1)$, denoted as
$\phi(\phi_0,\phi_1)$, since the power control policies $\mu$
depend on $(\phi_0,\phi_1)$. Now that $\phi_0$ and $\phi_1$ take
values from $[0,1]$ and the RHS function takes values from
$[0,1]$, we can find the solution through an iterative algorithm
according to the fixed-point theorem\footnote{Although trivially, either
$\phi_0$ or $\phi_1$ taking value 0 or 1 will at the same time turn
out to be some value in (0,1) for the function
$\phi(\phi_0,\phi_1)$, which tells us that the solution of $\phi_0$
or $\phi_1$ is in (0,1).}.

It is clear that $(\mu_0,\mu_1)$ are the solutions to the following
\begin{align}
&\frac{\lambda_0}{\phi_0\log_e2}(1+\mu_0z_M)^{-\beta-1}z_M-\kappa=0\label{eq:optcond1}\\
&\frac{\lambda_0}{\phi_0\log_e2}\left(1+\frac{\mu_0z_M}{1+\mu_1
z_M}\right)^{-\beta-1}\frac{z_M}{1+\mu_1z_M}-\kappa=0\label{eq:optcond2}\\
&-\frac{\lambda_0}{\phi_0\log_e2}\left(1+\frac{\mu_0 z_M}{1+\mu_1
z_M}\right)^{-\beta-1}\frac{\mu_0z_M^2}{(1+\mu_1
z_M)^2}\nonumber\\
&\hspace{.5cm}+\frac{\lambda_1}{\phi_1\log_e2}\left(\frac{1+\mu_1
z_M}{1+\gamma\mu_1z_E}\right)^{-\beta-1}\frac{z_M-\gamma
z_E}{(1+\gamma\mu_1z_E)^2}-\kappa=0\label{eq:optcond3}
\end{align}
where (\ref{eq:optcond1})-(\ref{eq:optcond3}) are obtained by taking
the derivative of $\mathcal{J}$ with respect to $\mu_0$ for $\mathbf{z} \in
\mathcal{Z}^c$, $\mu_0$ for $\mathbf{z} \in \mathcal{Z} $, and
$\mu_1$ for $\mathbf{z} \in \mathcal{Z}$, respectively. Whenever
$\mu_0$ and $\mu_1$ turn out to have negative values through these
equations, they are set to 0 according to the convexity of the
optimization problem.
Although the closed form expressions for $(\mu_0,\mu_1)$ are hard to
find, we can get some insight by examining
(\ref{eq:optcond1})-(\ref{eq:optcond3}). Define
$\alpha_1=\frac{\kappa\phi_0\log_e2}{\lambda_0}$ and
$\alpha_2=\frac{\kappa\phi_1\log_e2}{\lambda_1}$, which are
chosen to satisfy the average power constraint (\ref{eq:avgpower})
with equality. Not surprisingly, $\mu_0$ in $\mathcal{Z}^c$ behaves
similarly as in point-to-point transmission \cite{jia}. Now, consider
(\ref{eq:optcond3}). In order for $\mu_1$ to have a nonnegative value,
the following should be satisfied
\begin{equation}\label{eq:subscond1}
\frac{z_M-\gamma z_E}{\alpha_2}-1\ge0.
\end{equation}
If $\mu_0=0$, we have from (\ref{eq:optcond3}) that
\begin{equation}\label{eq:subscond2}
\frac{1}{\alpha_2}\left(\frac{1+\mu_1
z_M}{1+\gamma\mu_1z_E}\right)^{-\beta-1}\frac{z_M-\gamma
z_E}{(1+\gamma\mu_1z_E)^2}-1=0.
\end{equation}
After a simple computation using(\ref{eq:optcond2}), we get
\begin{equation}\label{eq:subscond3}
\frac{\mu_0z_M}{1+\mu_1z_M}=\left(\frac{z_M}{\alpha_1(1+\mu_1z_M)}\right)^{\frac{1}{\beta+1}}-1
\end{equation}
which gives us that $\mu_0<0$ if
\begin{equation}\label{eq:subscond4}
\frac{z_M}{\alpha_1(1+\mu_1 z_M)}<1.
\end{equation}
This tells us that when the power allocated to confidential message
is large enough, there should be no common message transmission.
Furthermore, substituting (\ref{eq:optcond2}) and (\ref{eq:subscond2})
into (\ref{eq:optcond3}), we have
\begin{small}
\begin{equation}\label{eq:subscond5}
\frac{1}{\alpha_2}\left(\frac{1+\mu_1z_M}{1+\gamma\mu_1
z_E}\right)^{-\beta-1}\frac{z_M-\gamma z_E}{(1+\mu_1
z_E)^2}-\left(\frac{z_M}{\alpha_1(1+\mu_1z_M)}\right)^{\frac{1}{\beta+1}}=0
\end{equation}
\end{small}
This is for the case when both $\mu_0$ and $\mu_1$ turn out to have
positive values directly from the optimization condition equations.

The following algorithm can be used to determine the optimal power
allocation:
\begin{codebox}
\Procname{$\proc{Algorithm\ PCI }(Power\ Control\ I)$} \li Given
$\lambda_0,\lambda_1$, obtain $\kappa^*,\phi_0^*,\phi_1^*$; \li
Denote
 $\alpha_1=\frac{\kappa^*\phi_0^*\log_e2}{\lambda_0}$,
 $\alpha_2=\frac{\kappa^*\phi_1^*\log_e2}{\lambda_1}$; \li \If $z_M-\gamma z_E>\alpha_2$ \li \Then Compute $\mu_1$
from (\ref{eq:subscond2}); \li \If
$\mu_1>\frac{1}{\alpha_1}-\frac{1}{z_M}$ or $z_M<\alpha_1$ \li \Then
$\mu_0=0$; \li \Else \If $\frac{z_M-\gamma
z_E}{\alpha_2}>\left(\frac{z_M}{\alpha_1}\right)^{\frac{1}{\beta+1}}$
\li \Then Compute $\mu_1$ from (\ref{eq:subscond5});\li Substitute
$\mu_1$ to (\ref{eq:optcond2}) to get $\mu_0$;\End \li \Else
$\mu_1=0$,
 $\mu_0=\left[\frac{1}{\alpha_1^{\frac{1}{\beta+1}}z_M^{\frac{\beta}{\beta+1}}}-\frac{1}{z_M}\right]^+$;\End
 \li \Else $\mu_1=0$,
 $\mu_0=\left[\frac{1}{\alpha_1^{\frac{1}{\beta+1}}z_M^{\frac{\beta}{\beta+1}}}-\frac{1}{z_M}\right]^+$;
 \End
\end{codebox}
The optimal values $\kappa^*,\phi_0^*,\phi_1^*$ can be numerically
computed to satisfy the average power constraint.

With the above analysis of the optimal power allocation policy of
$(\mu_0,\mu_1)$, the condition (\ref{eq:case1cond}) essentially
requires that the power allocated to $\mu_1$ is large such that the
interference from sending confidential messages is stronger in the
expression for $R_{01}(\mu)$, i.e., higher effective secrecy
throughput.

\textbf{Case II: } The maximal instantaneous common message rate is
given by the rate of receiver 2
\begin{align}
R_{02}=\left\{\begin{array}{ll}
\log_2\left(1+\frac{\gamma\mu_0(\mathbf{z})z_E}{1+\gamma\mu_1(\mathbf{z})z_E}\right),\,&\mathbf{z}\in\mathcal{Z}\\
\log_2\left(1+\gamma\mu_0(\mathbf{z})z_E\right),\,&\mathbf{z}\in\mathcal{Z}^c\end{array}\right.
\end{align}
as long as the obtained power control policy satisfies
\begin{equation}\label{eq:case2cond}
R_{01}(\mu)>R_{02}(\mu).
\end{equation}
Then, the Lagrangian is given by
\begin{small}
\begin{align}
\mathcal{J}&=-\frac{\lambda_0}{\beta \log_e 2}\log_e
\Bigg(\int_{\mathbf{z}\in\mathcal{Z}}\left(1+\frac{\gamma\mu_0(\mathbf{z})z_E}{1+\gamma\mu_1(\mathbf{z})
z_E}\right)^{-\beta}p_{\mathbf{z}}(z_M,z_E)d\mathbf{z}\nonumber\\
&\hspace{1.5cm}+\int_{\mathbf{z}\in\mathcal{Z}^c}\left(1+\gamma\mu_0(\mathbf{z})z_E\right)^{-\beta}p_{\mathbf{z}}(z_M,z_E)d\mathbf{z}\Bigg)\nonumber\\
&-\frac{\lambda_1}{\beta \log_e
2}\log_e\Bigg(\int_{\mathbf{z}\in\mathcal{Z}}\left(\frac{1+\mu_1(\mathbf{z})z_M}{1+\gamma\mu_1(\mathbf{z})
z_M}\right)^{-\beta}p_{\mathbf{z}}(z_M,z_E)d\mathbf{z}\nonumber\\
&\hspace{1.5cm}+\int_{\mathbf{z}\in\mathcal{Z}^c}p_{\mathbf{z}}(z_M,z_E)d\mathbf{z}\Bigg)\nonumber\\
&\hspace{1cm}-\kappa\left(\E_{\mathbf{z}\in\mathcal{Z}
}\{\mu_0(\mathbf{z})+\mu_1(\mathbf{z})\}+\E_{\mathbf{z}\in\mathcal{Z}^c}\{\mu_0(\mathbf{z})\}\right)
\end{align}
\end{small}
$(\mu_0,\mu_1)$ are the solutions to the following
\begin{small}
\begin{align}
&\frac{\lambda_0}{\phi_0\log_e2}(1+\gamma\mu_0z_E)^{-\beta-1}\gamma z_E-\kappa=0\label{eq:2optcond1}\\
&\frac{\lambda_0}{\phi_0\log_e2}\left(1+\frac{\gamma\mu_0z_E}{1+\gamma\mu_1
z_E}\right)^{-\beta-1}\frac{\gamma z_E}{1+\gamma\mu_1z_E}-\kappa=0\label{eq:2optcond2}\\
&-\frac{\lambda_0}{\phi_0\log_e2}\left(1+\frac{\gamma\mu_0
z_E}{1+\gamma\mu_1 z_E}\right)^{-\beta-1}\frac{\mu_0(\gamma
z_E)^2}{(1+\gamma\mu_1
z_E)^2}\nonumber\\
&\hspace{.5cm}+\frac{\lambda_1}{\phi_1\log_e2}\left(\frac{1+\mu_1
z_M}{1+\gamma\mu_1z_E}\right)^{-\beta-1}\frac{z_M-\gamma
z_E}{(1+\gamma\mu_1z_E)^2}-\kappa=0\label{eq:2optcond3}
\end{align}
\end{small}
where (\ref{eq:2optcond1})-(\ref{eq:2optcond3}) are obtained by
taking the derivative of $\mathcal{J}$ with respect to $\mu_0$ for $\mathbf{z}
\in \mathcal{Z}^c$, $\mu_0$ for $\mathbf{z} \in \mathcal{Z} $, and
$\mu_1$ for $\mathbf{z} \in \mathcal{Z}$, respectively. Whenever
$\mu_0$ or $\mu_1$ turns out to have negative values through these
equations, they are again set to 0.

Following a similar analysis as shown for \textbf{Case I}, we obtain the
following algorithm to determine the optimal power control policies.
\begin{codebox}
\Procname{$\proc{Algorithm\ PC-II }(Power\ Control\ II)$} \li Given
$\lambda_0,\lambda_1$, obtain $\kappa^*,\phi_0^*,\phi_1^*$; \li
Denote
 $\alpha_1=\frac{\kappa^*\phi_0^*\log_e2}{\lambda_0}$,
 $\alpha_2=\frac{\kappa^*\phi_1^*\log_e2}{\lambda_1}$; \li \If $z_M-\gamma z_E>\alpha_2$ \li \Then Compute $\mu_1$
from (\ref{eq:subscond2}); \li \If
$\mu_1>\frac{1}{\alpha_1}-\frac{1}{\gamma z_E}$ or $\gamma
z_E<\alpha_1$ \li \Then $\mu_0=0$; \End\li \Else \If
$\frac{z_M-\gamma z_E}{\alpha_2}>\left(\frac{\gamma
z_E}{\alpha_1}\right)^{\frac{1}{\beta+1}}$ \li \Then Compute $\mu_0$
and $\mu_1$ from (\ref{eq:2optcond2}) and (\ref{eq:2optcond3});\End
\li \Else $\mu_1=0$,
 $\mu_0=\left[\frac{1}{\alpha_1^{\frac{1}{\beta+1}}(\gamma z_E)^{\frac{\beta}{\beta+1}}}-\frac{1}{\gamma z_E}\right]^+$;\End
 \li \Else $\mu_1=0$,
 $\mu_0=\left[\frac{1}{\alpha_1^{\frac{1}{\beta+1}}(\gamma z_E)^{\frac{\beta}{\beta+1}}}-\frac{1}{\gamma z_E}\right]^+$;
 \End
\end{codebox}
where $\kappa^*,\phi_0^*,\phi_1^*$ can be numerically computed to
satisfy the average power constraint.

According the above optimal power allocation policy of
$(\mu_0,\mu_1)$, in contrast to \textbf{Case I}, the condition
(\ref{eq:case2cond}) indicates that the power allocated to $\mu_1$
is small such that the interference from sending confidential
messages can be ignored in $R_{01}(\mu)$, i.e., smaller effective
secrecy throughput.

\textbf{Case III: } The first two sub-cases $A)$ and $B)$ are
trivial because, other than the condition $R_{01}(\mu)=R_{02}(\mu)$,
there is no difference in the power allocation policies from what we have derived
in \textbf{Case I} and \textbf{Case II}. We are more interested in
the case in which there is $0<\delta^*<1$ decided by the following
condition
\begin{equation}
R_{01}(\mu^{\delta^*})=R_{02}(\mu^{\delta^*}) \text{and
}\C_{01}(\mu^{\delta^*})=\C_{02}(\mu^{\delta^*}).
\end{equation}
We will first derive the optimal power control policies for any
given $\delta^*$, and then determine $\delta^*$.

The Lagrangian is given by
\begin{small}
\begin{align}
\hspace{-1.5cm}\mathcal{J}&=-\frac{\lambda_0}{\beta\log_e2}\Bigg[\delta
\log_e\Bigg(\int_{\mathbf{z}\in\mathcal{Z}}\left(1+\frac{\mu_0(\mathbf{z})z_M}{1+\mu_1(\mathbf{z})
z_M}\right)^{-\beta}p_{\mathbf{z}}(z_M,z_E)d\mathbf{z}\nonumber\\
&\hspace{1.5cm}+\int_{\mathbf{z}\in\mathcal{Z}^c}\left(1+\mu_0(\mathbf{z})z_M\right)^{-\beta}p_{\mathbf{z}}(z_M,z_E)d\mathbf{z}\Bigg)\nonumber\\
&+(1-\delta)\log_e
\Bigg(\int_{\mathbf{z}\in\mathcal{Z}}\left(1+\frac{\gamma\mu_0(\mathbf{z})z_E}{1+\gamma\mu_1(\mathbf{z})
z_E}\right)^{-\beta}p_{\mathbf{z}}(z_M,z_E)d\mathbf{z}\nonumber\\
&\hspace{1.5cm}+\int_{\mathbf{z}\in\mathcal{Z}^c}\left(1+\gamma\mu_0(\mathbf{z})z_E\right)^{-\beta}p_{\mathbf{z}}(z_M,z_E)d\mathbf{z}\Bigg)\Bigg]\nonumber\\
&-\frac{\lambda_1}{\beta \log_e
2}\log_e\Bigg(\int_{\mathbf{z}\in\mathcal{Z}}\left(\frac{1+\mu_1(\mathbf{z})z_M}{1+\gamma\mu_1(\mathbf{z})
z_M}\right)^{-\beta}p_{\mathbf{z}}(z_M,z_E)d\mathbf{z}\nonumber\\
&\hspace{1.5cm}+\int_{\mathbf{z}\in\mathcal{Z}^c}p_{\mathbf{z}}(z_M,z_E)d\mathbf{z}\Bigg)\nonumber\\
&\hspace{1cm}-\kappa\left(\E_{\mathbf{z}\in\mathcal{Z}
}\{\mu_0(\mathbf{z})+\mu_1(\mathbf{z})\}+\E_{\mathbf{z}\in\mathcal{Z}^c}\{\mu_0(\mathbf{z})\}\right)
\end{align}
\end{small}
$(\mu_0,\mu_1)$ are the solutions to the following
\begin{small}
\begin{align}
&\hspace{-1cm}\frac{\lambda_0}{\phi_0\log_e2}\left(\delta(1+\mu_0z_M)^{-\beta-1}z_M+(1-\delta)(1+\gamma\mu_0z_E)^{-\beta-1}\gamma z_E\right)-\kappa=0\label{eq:3optcond1}\\
&\frac{\lambda_0}{\phi_0\log_e2}\Bigg(\delta\left(1+\frac{\mu_0z_M}{1+\mu_1
z_M}\right)^{-\beta-1}\frac{z_M}{1+\mu_1z_M}\nonumber\\
&+(1-\delta)\left(1+\frac{\gamma\mu_0z_E}{1+\gamma\mu_1
z_E}\right)^{-\beta-1}\frac{\gamma z_E}{1+\gamma\mu_1z_E}\Bigg)-\kappa=0\label{eq:3optcond2}\\
&-\frac{\lambda_0}{\phi_0\log_e2}\Bigg(\delta\left(1+\frac{\mu_0
z_M}{1+\mu_1 z_M}\right)^{-\beta-1}\frac{\mu_0z_M^2}{(1+\mu_1
z_M)^2}\nonumber\\
&+(1-\delta)\left(1+\frac{\gamma\mu_0 z_E}{1+\gamma\mu_1
z_E}\right)^{-\beta-1}\frac{\mu_0(\gamma z_E)^2}{(1+\gamma\mu_1
z_E)^2}\Bigg)\nonumber\\
&\hspace{.5cm}+\frac{\lambda_1}{\phi_1\log_e2}\left(\frac{1+\mu_1
z_M}{1+\gamma\mu_1z_E}\right)^{-\beta-1}\frac{z_M-\gamma
z_E}{(1+\gamma\mu_1z_E)^2}-\kappa=0\label{eq:3optcond3}
\end{align}
\end{small}
where (\ref{eq:2optcond1})-(\ref{eq:2optcond3}) are obtained by
taking the derivative of $\mathcal{J}$ with respect to $\mu_0$ for $\mathbf{z}
\in \mathcal{Z}^c$, $\mu_0$ for $\mathbf{z} \in \mathcal{Z} $, and
$\mu_1$ for $\mathbf{z} \in \mathcal{Z}$, respectively. Similarly as before, whenever
$\mu_0$ or $\mu_1$ have negative values through these
equations, they are set to 0.
Considering (\ref{eq:3optcond2}), we see that when $\mu_0=0$, $\mu_1$ needs to
satisfy
\begin{small}
\begin{equation}\label{eq:3subcond1}
\frac{\lambda_0}{\phi_0\log_e2}\left(\frac{\delta
z_M}{1+\mu_1z_M}+\frac{(1-\delta)\gamma
z_E}{1+\gamma\mu_1z_E}\right)-\kappa\le0
\end{equation}
\end{small}
and $\mu_1$ is given by (\ref{eq:3optcond3})
\begin{small}
\begin{equation}\label{eq:3subcond2}
\frac{\lambda_1}{\phi_1\log_e2}\left(\frac{1+\mu_1
z_M}{1+\gamma\mu_1z_E}\right)^{-\beta-1}\frac{z_M-\gamma
z_E}{(1+\gamma\mu_1z_E)^2}-\kappa=0
\end{equation}
\end{small}

When $\mu_1=0$, $\mu_0$ is given by
\begin{small}
\begin{align}\label{eq:3subcond3}
&\frac{\lambda_0}{\phi_0\log_e2}\Bigg(\delta\left(1+\mu_0z_M\right)^{-\beta-1}z_M
+\nonumber\\
&(1-\delta)\left(1+\gamma\mu_0z_E\Bigg)^{-\beta-1}\gamma
z_E\right)-\kappa=0
\end{align}
\end{small}

Now, for $\mu_1\ge0$, we need to have the following
\begin{small}
\begin{align}\label{eq:3subcond4}
&-\frac{\lambda_0}{\phi_0\log_e2}\Bigg(\delta\left(1+\mu_0
z_M\right)^{-\beta-1}\mu_0z_M^2+\nonumber\\
&(1-\delta)\left(1+\gamma\mu_0 z_E\right)^{-\beta-1}\mu_0(\gamma
z_E)^2\Bigg)+\frac{\lambda_1(z_M-\gamma
z_E)}{\phi_1\log_e2}-\kappa\ge0
\end{align}
\end{small}
where $\mu_0$ is computed from (\ref{eq:3subcond3}).

For any $\delta$, we need to find the associated power control
policy $(\mu_0,\mu_1)$ satisfying
(\ref{eq:3optcond1})-(\ref{eq:3optcond3}). Then, we need to further
search over $0<\delta<1$ for $\delta^*$ that satisfies
\begin{equation}
\C_{01}(\mu)=\C_{02}(\mu).
\end{equation}

We obtain the following algorithm to determine the optimal power
control policies.
\begin{codebox}
\Procname{$\proc{Algorithm\ PC-III-C }(Power\ Control\ III-C)$} \li
Given $\lambda_0,\lambda_1$, obtain $\kappa^*,\phi_0^*,\phi_1^*$;
\li Denote
 $\alpha_1=\frac{\kappa^*\phi_0^*\log_e2}{\lambda_0}$,
 $\alpha_2=\frac{\kappa^*\phi_1^*\log_e2}{\lambda_1}$; \li \If $z_M-\gamma z_E>\alpha_2$ \li \Then Compute $\mu_1$
from (\ref{eq:3subcond2}); \li \If (\ref{eq:3subcond1}) holds or
$\delta z_M +(1-\delta)\gamma z_E<\alpha_1$; \li\Then
$\mu_0=0$;\End\li \Else \If (\ref{eq:3subcond4}) holds \li \Then
Compute $\mu_0$ and $\mu_1$ from (\ref{eq:3optcond2}) and
(\ref{eq:3optcond3});\End \li \Else $\mu_1=0$,
 $\mu_0$ is given by (\ref{eq:3subcond3});\End
 \li \Else $\mu_1=0$,
 $\mu_0$ is given by (\ref{eq:3subcond3});
 \End
\end{codebox}
where $\kappa^*,\phi_0^*,\phi_1^*$ can be numerically computed to
satisfy the average power constraint.

Based on the previous results, we have the following algorithm to
find the optimal power control policies.
\begin{codebox}
\Procname{$\proc{Algorithm\ PC }(Power\ Control)$} \li Find
$\mu^{(1)}$ given in \textbf{Case I}; \li \If
$R_{01}(\mu^{(1)})<R_{02}(\mu^{(1)})$\li \Then $\mu^*=\mu^{(1)}$;
\li\Else \If $R_{01}(\mu^{(1)})=R_{02}(\mu^{(1)})$ and
$\C_{01}(\mu)>\C_{02}(\mu)$ \li \Then $\mu^*=\mu^{(1)}$;\End\li
\Else Find $\mu^{(2)}$ given in \textbf{Case II}; \li\If
$R_{01}(\mu^{(2)})>R_{02}(\mu^{(2)})$ \li \Then
$\mu^*=\mu^{(2)}$;\End\li\Else \If
$R_{01}(\mu^{(2)})=R_{02}(\mu^{(2)})$ and
$\C_{01}(\mu)<\C_{02}(\mu)$\li\Then $\mu^*=\mu^{(2)}$;\End \li\Else
For a given $\delta$, find $\mu^{(3)}$ given in \textbf{Case
III-C};\li Search over $0\le\delta\le1$ to find $\delta$ that
satisfies \li $R_{01} (\mu^{(3)})= R_{02}(\mu^{(3)})$ and $
\C_{01}(\mu^{(3)})=\C_{02}(\mu^{(3)})$ \li $\mu^*=\mu^{(3)}$. \End
\end{codebox}

\begin{figure}
\begin{center}
\includegraphics[width=\figsize\textwidth]{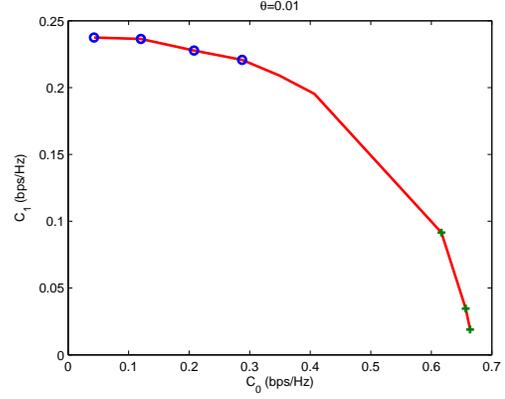}
\caption{The achievable secrecy throughput region. $T=2$ ms,
$B=10^5$ Hz, and $\tsnr=0$ dB.}\label{fig:region1}
\end{center}
\end{figure}

In Fig. \ref{fig:region1}, we plot the achievable effective
secrecy throughput region in Rayleigh fading channel. We assume that
$\gamma=1$, i.e., the noise variances at both receivers are equal. In the figure, the circles
fall into Case I or Case III-A, and the pluses fall into Case II or Case
III-B, and Case III-C is shown as line only.
\section{Conclusion}
In this paper, we have investigated the fading broadcast channels
with confidential message under statistical QoS constraints. We have
first defined the effective secrecy throughput
region, which was later proved to be convex. Then, the problem of
finding points on the boundary of the throughput region is shown to be equivalent
to solving a series of optimization problem. We have extended the
approach used in previous studies to the scenario considered in this
paper. Following similar steps, we have determined the conditions
satisfied by the optimal power control policies. In particular, we have
identified the algorithms for computing the power allocated to each
fading state from the optimality conditions. Numerical results are
provided as well.

%

\end{document}